# Role of inclusiveness of learning environment in predicting students' outcomes in courses in which women are not underrepresented

Sonja Cwik
University of Pittsburgh

Chandralekha Singh
University of Pittsburgh

*Student beliefs in introductory physics courses can influence their course outcomes and retention in STEM disciplines and future career aspirations. This study used survey data from 501 students in the first of two-semester algebra-based introductory physics courses primarily taken by bioscience majors, in which women make up approximately 65% of the class. We investigated how the learning environment including perceived recognition, peer interaction, and sense of belonging students' correlate with physics outcomes, including their physics self-efficacy, interest, and identity. We found that in general, women had lower physics beliefs than men and the learning environment plays a major role in explaining student outcomes. We also found that perceived recognition played an important role in predicting students' physics identity and students' sense of belonging played an important role in predicting students' physics self-efficacy in the first algebra-based introductory physics course investigated. These findings can be useful to contemplate strategies to create an equitable and inclusive learning environment to help all students to excel in these physics courses.*

*Equity and inclusion, gender, introductory physics courses, motivational beliefs*

**INTRODUCTION AND THEORETICAL FRAMEWORK**

In the past few decades, there has been a focus on the experiences and participation of women in many science, technology, engineering, and math (STEM) fields (Blue et al., 2018; Buck, 2002; Center, 2015; Salmi & Thuneberg, 2019; Seymour & Hewitt, 1997; Whitten et al., 2003). Some research studies have focused on students' beliefs in different STEM domains which can influence students' continuation in related courses, majors, and careers (Beilock et al., 2007; Eccles, 1994; Hewitt & Seymour, 1992; Maries et al., 2018; Marshman, Kalender, Nokes-Malach, et al., 2018; Nokes-Malach et al., 2018; Tobias, 1990). Some of these studies that focus on women and ethnic and racial minority students in physics show that they, in general, have lower beliefs and grades than men (Archer et al., 2017; Lock et al., 2013; Monsalve et al., 2016). Inequitable outcomes like this may be a result of inequitable access to resources, inadequate support, and inequitable learning environments.

Our conceptualization of equity in learning includes three pillars: equitable opportunities to learn, equitable and inclusive learning environments, and equitable outcomes. Equity in learning would require all students to have equitable opportunities and access to resources and that students have an equitable and inclusive learning environment with appropriate support and mentoring so that they can engage in learning in a meaningful and enjoyable manner. Equitable learning outcome means that students from all demographic groups (e.g., regardless of their gender identity, etc.) who have the pre-requisites to enroll in courses have comparable learning outcomes. This conceptualization of equitable outcome is consistent with Rodriguez et. al.'s equity of parity model (Rodriguez et al., 2012). We note that equitable and inclusive learning environments and equitable outcomes are intricately tied to each other.

The formation and development of students' science identity have been a focus of many prior studies (Andersen et al., 2014; Dou & Cian, 2020; Hazari et al., 2010; Hughes et al., 2013; Kalender et al., 2019a; Vincent-Ruz & Schunn, 2019). Science identity is defined as identifying with academic domains in science, i.e., whether students see themselves as science people (Chen & Wei, 2020; Hazari et al., 2010; Monsalve et al., 2016) or those who can excel in science. Students' identity in STEM disciplines has been shown to play an important role in their in-class participation and choices of majors and careers (Carlone & Johnson, 2007; Gee, 2000; Krogh & Andersen, 2013; Stets et al., 2017; Vincent-Ruz & Schunn, 2018). Studies have shown that it can be more difficult for women to form a physics identity than men (Archer et al., 2017; Godwin et al., 2016; Lock et al., 2013; Monsalve et al., 2016). However, most of the studies in the college context concerning physics identity and factors that influence it are conducted in classes in which women are underrepresented (Hazari et al., 2010; Kalender et al., 2019a). Students' identity in a domain can be context dependent (Gee, 2000), therefore, it is important to examine the role of the inclusiveness of the learning environment on the physics identity of students in courses in which women outnumber men, e.g., in introductory physics courses for bioscience majors.

When examining equity in physics education and the role of inclusiveness of the learning environment, it is useful to examine the practices that lead to inequities (Ladson-Billings & Tate IV, 1995; Rios-Aguilar, 2014; Yosso, 2005). Physics is a discipline with problematic stereotypes and biases about who belongs in it and can excel in it. One common stereotype is that only high achievers or geniuses can excel in physics (Leslie et al., 2015). However, genius is more often associated with boys (Upson & Friedman, 2012), and girls from a young age tend to shy away from fields associated with innate brilliance or genius (Bian et al., 2017). These stereotypes can continue to impact women as they get older. Teachers and school counselors pay more attention to male students and counselors give gendered advice to students regarding high school physics and math courses to take and majors to pursue when in college. These stereotypes and biases are also prevalent at the university level. One study found that biology and physics faculty members rated a male student as significantly more competent than a female student when presented with a hypothetical scenario for hiring a student for lab work with either a male or female name (Moss-Racusin et al., 2012). These highly problematic stereotypes and biases are founded in the historical marginalization of certain groups, e.g., women in physics, and continue to manifest today in many ways including, gendered beliefs and barriers to women excelling in physics when there is no explicit focus on making the learning environment equitable and inclusive. The pervasive stereotypes and biases can impact women's identity even in these physics courses in which they are not underrepresented in an inequitable and non-inclusive environment.

Since science identity is key to students' success, it is important to investigate students' beliefs that play a role in identity formation. Carlone and Johnson's science identity framework includes three interrelated dimensions: competence, performance, and recognition by others (Johnson et al., 2017). However, in a study of introductory physics students by Hazari et al., performance and competence were found to be highly correlated and a fourth dimension, interest was added to the model (Hazari et al., 2010). In a slightly reframed version of the physics identity framework (Kalender et al., 2019b), students' self-efficacy, interest, and perceived recognition by others have been shown to predict students' physics identity (Cwik & Singh, 2022b; Li & Singh, 2022a).

Self-efficacy in a particular discipline is the students' belief in their ability to succeed in a particular task or course (Bandura, 1994). It has been shown to impact students' engagement, learning, and persistence in science courses (Bouffard-Bouchard et al., 1991; Cavallo et al., 2004; Fencl & Scheel, 2005; Larry & Wendt, 2021; Lindstrøm & Sharma, 2011; Nissen & Shemwell, 2016; Sawtelle et al., 2012; Schunk & Pajares, 2002). Students with high self-efficacy in a domain are less likely to have anxiety that can rob them of their cognitive resources while learning and test taking, since the working memory during problem solving has limited capacity (Maloney et al., 2014). They are also more likely to engage in effective learning strategies and are less likely to procrastinate while engaging with learning tools. Thus, self-efficacy can be a predictor of student performance in that domain (Zimmerman, 2000). However, in physics courses, gender differences have been shown in students' self-efficacy.(Cwik & Singh, 2021a, 2022a; Kalender et al., 2020)

Similarly, interest in a particular discipline can affect students' perseverance and achievement in that discipline (Hidi, 2006; Serrano Corkin et al., 2021; Strenta et al., 1994; Wang & Degol, 2013). One study showed that changing the curriculum to stimulate the interest of girls helped improve all the students' understanding at the end of the year (Häussler & Hoffmann, 2002). Furthermore, according to expectancy-value theory, self-efficacy and interest are related constructs that predict student outcomes and career expectations (Wigfield & Eccles, 1992). Moreover, students' interest and self-efficacy can be connected to their interaction with and recognition by other people (Bandura, 1991; Hidi, 2006).

A student's perceived recognition by others has also been shown to play an important role in a student's identity (Vincent-Ruz & Schunn, 2018) and is a particularly important factor that influences women's other beliefs (Goodenow, 1993). Studies have shown that female students do not feel recognized appropriately even before they enter college (Archer et al., 2017; Bian et al., 2017; Kalender et al., 2019a). Our prior individual interviews suggest that students' perceived recognition by instructors and teaching assistants impacts their self-efficacy and interest in physics, e.g., see (Doucette et al., 2020; Doucette & Singh, 2020). In addition, other learning environment factors, such as students' sense of belonging and perception of the effectiveness of interaction with their peers are important. Students' sense of belonging in physics has been shown to correlate with their retention and self-efficacy (Goodenow, 1993; Masika & Jones, 2016) and students' interaction with peers has been shown to enhance understanding and engagement in courses (Meltzer & Manivannan, 2002). These factors have been shown to correlate with students' physics grades (Cwik & Singh, 2021b, 2022c, 2022d; Li et al., 2020). In addition, students' self-efficacy, interest, and students' perception of the learning environment has been shown to correlate with students' physics beliefs and engineering identity in calculus-based physics courses where men outnumber women (Li & Singh, 2021, 2022b, 2022c). Since these factors can impact students' physics beliefs, it is important to investigate student perceptions of the inclusiveness of the learning environment and how they influence the outcomes pertaining to physics self-efficacy, interest, and identity in the algebra-based physics courses where women outnumber men.

This study examined students' perception of the inclusiveness of learning environment (including their sense of belonging, perceived recognition, and peer interaction) on their self-efficacy, interest, and identity at the end of an algebra-based introductory physics course controlling for students' gender, self-efficacy, and interest at the beginning of the course (Fig. 1). Student perceptions of the inclusiveness of the learning environment are not only formed by classroom experiences but also experiences they have outside of the classroom. For example, the perceptions of the inclusiveness of the learning environment could also be shaped by students asking for help during office hours, their email correspondence with the instructor or teaching assistant (TA), and students studying together. Therefore, the perception of the inclusiveness of learning environment factors in our investigation include students' perceptions of their interaction with their peers (from now on simply referred to as peer interaction for brevity), sense of belonging (referred to as belonging for brevity), and perceived recognition by others (including friends, family, and their instructors/TAs). We control for students' gender, self-efficacy, and interest at the beginning of the course, which are related to students' beliefs about physics before they enter the classroom. A schematic diagram of our model with only the path analysis is shown in Fig 1.

We first delineate the research questions based upon our framework, then describe the methodology, show the results and discussion, and finally conclude with instructional implications and future directions. We use structural equation modeling (SEM) (*R: A Language and Environment for Statistical Computing*, 2019) to investigate how the perception of the inclusiveness of the learning environment predicts students' physics self-efficacy, interest, and identity at the end of the course. The following research questions were answered by analyzing data from a validated survey administered to students in the first semester of algebra-based physics courses (physics 1) primarily for bioscience majors at a large public research university in the US in which women outnumber men and using mediation analysis via SEM:

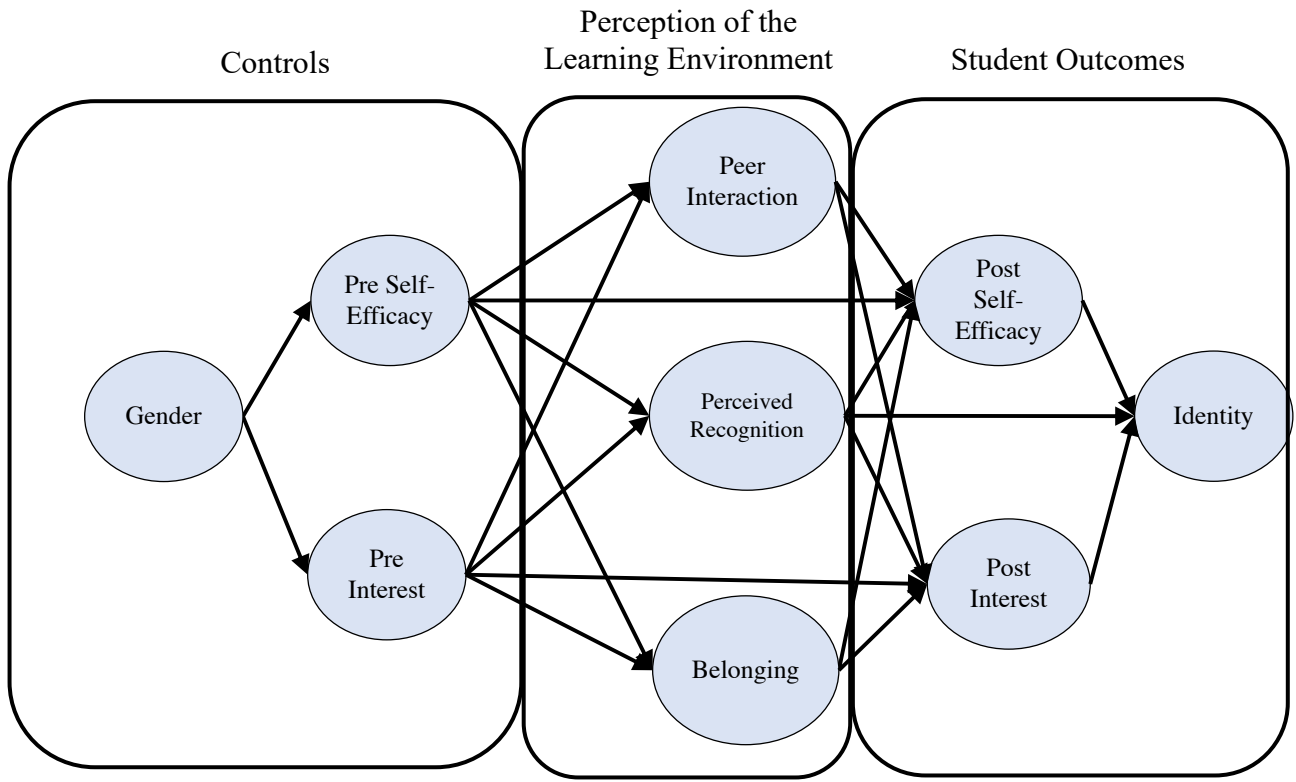

**FIG. 1 ALL REGRESSION PATHS WERE CONSIDERED FROM LEFT TO RIGHT IN OUR MODEL; HOWEVER, ONLY SOME OF THE PATHS ARE SHOWN FOR CLARITY.**

**RQ1** Are there gender differences in students' physics beliefs (self-efficacy, interest, and identity) at the end of the course and do interest and self-efficacy change from the beginning to the end of the course?

**RQ2** Is there moderation by gender for any of the regression paths?

**RQ3** If moderation does not affect any path, does gender mediate any of the controlled factors, perceived inclusiveness of learning environment factors, or student outcomes?

**RQ4** How does the inclusiveness of the learning environment (including perceived recognition, belonging, and peer interaction) predict physics identity, physics self-efficacy, and physics interest at the end of the course?

**METHODOLOGY**

**Participants**

We administered a validated written survey at a large public research university in the U.S. to students at the beginning (pre) and end (post) of the first semester of an algebra-based introductory physics course (called physics 1). The class is typically taken as a mandatory requirement by students on the pre-medical or pre-health track (mainly biosciences majors) primarily in their junior or senior year of undergraduate studies. The university provided demographic information such as age, gender, and ethnic/racial information using an honest broker process by which the research team received the information without knowledge of the identities of the participants. We analyzed data from 501 students who completed the pre/post survey on paper scantrons. From the university data, the participants were 35% male and 65% female students. We acknowledge that gender is not a binary construct. However, the data provided by the university only includes options of male and female and thus gender was used as a binary variable in our analysis. Less than 1% of the students did not provide this information and were not included in the analysis.

**Instrument Validity**

The items in the validated survey focused on different aspects of students' beliefs at the time the survey was administered (beginning and end of the course) that included their perception of the inclusiveness of the learning environment. In particular, the study focused on students' responses to validated survey items on physics identity, self-efficacy, interest, sense of belonging, perceived recognition, and peer interaction. The survey items were adapted from previously validated surveys (Glynn et al., 2011; Hammer, 1994; *PERTS Academic Mindsets Assessment*, 2020) and re-validated in our own context using one-on-one student interviews, Exploratory Factor Analysis (EFA), Confirmatory Factor Analysis (CFA) (Doucette & Singh, 2020; Marshman, Kalender, Schunn, et al., 2018), and Pearson correlations. Interviews with students suggest that they interpreted the items correctly. The CFA established a measurement model for the constructs and was used in SEM. In the CFA, the model fit indices were good and all the factor loadings were above 0.50, which indicates good loadings (Cohen, 2013). The results of the CFA model are shown in Table 1.

The *physics identity* items focus on whether the students see themselves as a physics person (Hazari et al., 2013). The *physics self-efficacy* items measure students' confidence in their ability to solve physics problems (Glynn et al., 2011; Hazari et al., 2013; Learning Activation Lab, 2017; Schell & Lukoff, 2010). The *interest in physics* items measure students' enthusiasm and curiosity to learn physics and ideas related to physics (Learning Activation Lab, 2017). The *sense of belonging* items evaluate whether students felt like they belonged in the introductory physics class (Goodenow, 1993; *PERTS Academic Mindsets Assessment*, 2020). The *perceived recognition* items measure the extent the student thought that other people see them as a physics person (Hazari et al., 2013). Lastly, the *peer interaction* items measure whether students thought that working with their peers was beneficial, e.g., in increasing their confidence to do physics (Sayer et al., 2016; Singh, 2005). The questions in the study were denoted on a Likert scale of 1 (low belief) to 4 (high belief) except for the sense of belonging questions which were designed on a scale of 1 to 5 to keep them consistent with the original survey (Likert, 1932).

**TABLE 1. SURVEY QUESTIONS FOR EACH OF THE CONSTRUCTS AND FACTOR LOADINGS FROM THE CONFIRMATORY FACTOR ANALYSIS (CFA) RESULT FOR ALL STUDENTS (N = 501). THE RATING SCALE FOR MOST OF THE SELF-EFFICACY AND INTEREST QUESTIONS WAS NO! NO YES YES! WHILE THE RATING SCALE FOR THE PHYSICS IDENTITY, PEER INTERACTION, AND PERCEIVED RECOGNITION QUESTIONS WERE: STRONGLY DISAGREE, DISAGREE, AGREE, STRONGLY AGREE. THE RATING SCALE FOR THE PHYSICS BELONGING QUESTIONS WAS: NOT AT ALL TRUE, A LITTLE TRUE, SOMEWHAT TRUE, MOSTLY TRUE, AND COMPLETELY TRUE. ALL P-VALUES ARE <0.001.**

| Construct and Item | Lambda |
|---|---|
| **Physics Identity** | |
| I see myself as a physics person. | 1.00 |
| **Physics Self-Efficacy** | |
| I am able to help my classmates with physics in the laboratory or recitation. | 0.60 |
| I understand concepts I have studied in physics. | 0.74 |
| If I study, I will do well on a physics test. | 0.76 |
| If I encounter a setback in a physics exam, I can overcome it. | 0.72 |
| **Physics Interest** | |
| I wonder about how physics works. | 0.57 |
| In general, I find physics. † | 0.75 |
| I want to know everything I can about physics. | 0.74 |
| I am curious about recent discoveries in physics. | 0.64 |
| **Physics Perceived Recognition** | |
| My family sees me as a physics person. | 0.89 |
| My friends see me as a physics person. | 0.92 |
| My physics instructor and/or TA sees me as a physics person. | 0.71 |
| **Physics Belonging** | |
| I feel like I belong in this class. | 0.81 |
| I feel like an outsider in this class. | 0.74 |
| I feel comfortable in this class. | 0.82 |
| I feel like I can be myself in this class. | 0.62 |
| Sometimes I worry that I do not belong in this class. | 0.69 |
| **Physics Peer Interaction** *My experiences and interactions with other students in this class…* | |
| made me feel more relaxed about learning physics. | 0.75 |
| increased my confidence in my ability to do physics. | 0.93 |
| increased my confidence that I can succeed in physics. | 0.96 |
| increased my confidence in my ability to handle difficult physics problems. | 0.90 |

† the rating scale for this question was very boring, boring, interesting, very interesting.

Pair-wise Pearson's *r* values indicate the correlation between each pair of constructs without controlling for the influence of any other factors (values are given in Table 2). Inter-correlations vary in strength, but none of the correlations are so high that the constructs cannot be separately examined, consistent with prior studies. In other words, all the correlations are low enough that all the constructs can be considered separate. Since the construct is the same at two different points of time and the course does not affect students' interest significantly, the highest inter-correlation was the value between pre-interest and post-interest (0.86). Our past work in calculus-based introductory physics courses has also shown that there is a high correlation in interest over the course of the introductory classes (Marshman, Kalender, Schunn, et al., 2018).

**TABLE 2. PEARSON INTER-CORRELATIONS ARE GIVEN FOR EACH PAIR OF CONSTRUCTS. ALL P-VALUES ARE < 0.001**

| Observed Variable | Pearson Correlation Coefficient | | | | | | | |
|---|---|---|---|---|---|---|---|---|
| | 1 | 2 | 3 | 4 | 5 | 6 | 7 | 8 |
| 1. Pre Self-Efficacy | -- | -- | -- | -- | -- | -- | -- | -- |
| 2. Pre Interest | 0.50 | -- | -- | -- | -- | -- | -- | -- |
| 3. Perceived Recognition | 0.41 | 0.41 | -- | -- | -- | -- | -- | -- |
| 4. Peer Interaction | 0.21 | 0.24 | 0.46 | -- | -- | -- | -- | -- |
| 5. Belonging | 0.48 | 0.32 | 0.59 | 0.62 | -- | -- | -- | -- |
| 6. Post Self-Efficacy | 0.54 | 0.35 | 0.61 | 0.62 | 0.81 | -- | -- | -- |
| 7. Post Interest | 0.34 | 0.86 | 0.61 | 0.47 | 0.51 | 0.60 | -- | -- |
| 8. Physics Identity | 0.40 | 0.45 | 0.78 | 0.43 | 0.58 | 0.60 | 0.62 | -- |

**Analysis**

We first analyzed descriptive statistics and compared female and male students' mean scores of the predictors and outcomes for statistical significance using *t*-tests and computed the effect size using Cohen's *d* (Cohen, 2013). The square of CFA factor loading (lambda) indicates the fraction of variance explained by the factor. For predictive relationships between different constructs, we used Structural Equation Modeling (SEM) as a statistical tool by using R (lavaan package) with a maximum likelihood estimation method (*R: A Language and Environment for Statistical Computing*, 2019). SEM is an extension of multiple regression and conducts several multiple regressions simultaneously between variables in one estimation model. This allows us to calculate the overall goodness of fit and allows for all estimates to be performed simultaneously so there can be a direct comparison between different structural components. We report model fit for SEM by using the Comparative Fit Index (CFI), Tucker-Lewis Index (TLI), Root Mean Square Error of Approximation (RMSEA), and Standardized Root Mean Square Residuals (SRMR). Commonly used thresholds for goodness of fit are as follows: CFI and TLI > 0.90, and RMSEA and SRMR < 0.08 (MacCallum et al., 1996).

The model estimates were performed using moderation analysis to check whether any of the relations between variables show differences across gender by using "lavaan" to conduct multi-group SEM. Initially, we tested different levels of measurement invariance model. In each step, we fixed different elements of the model to equality across gender and compared the results to the previous step using the Likelihood Ratio Test. Since we did not find significant moderation by gender, we tested the theoretical model in mediation analysis, using gender as a variable directly predicting items to examine the resulting structural paths between constructs.

**RESULTS AND DISCUSSION**

**Gender Differences in predictors and outcomes**

To answer **RQ1**, the mean values of all constructs in our model were statistically significantly different disadvantaging female students (TABLE 3). This pattern is qualitatively similar to what was previously observed in calculus-based courses (Kalender et al., 2019a, 2019b) even though women make up the majority of students in the algebra-based course (65%). Furthermore, for both self-efficacy (gray) and interest (blue), constructs that have scores from both the pre test and the post test, we find statistically significant differences in the scores from the beginning to the end of physics 1 for both women and men. One possible hypothesis for this disparity is the previous experiences and stereotypes about physics that women have internalized from an early age before starting college physics (both in K-12 education setting and outside of it interacting with family, friends, media, etc.). Additionally, Table 3 shows that women also have more negative experiences in the classroom, indicated by their lower scores on the perception of the inclusiveness of the learning environment constructs, despite women making up the majority of students in the class.

**TABLE 3 MEAN PREDICTOR AND OUTCOME VALUES AND EFFECT SIZES (COHEN'S D) BY GENDER. ALL *P*-VALUES ARE < 0.001.**

| Predictors and Outcomes (Score Range) | Mean | | Cohen's $d$ |
|---|---|---|---|
| | Male | Female | |
| Pre Self-Efficacy (1-4) | 3.07 | 2.83 | 0.58 |
| Pre Interest (1-4) | 2.74 | 2.45 | 0.58 |
| Perceived Recognition (1-4) | 2.22 | 1.92 | 0.43 |
| Peer Interaction (1-4) | 2.90 | 2.58 | 0.43 |
| Belonging (1-5) | 3.58 | 3.01 | 0.60 |
| Post Self-Efficacy (1-4) | 2.92 | 2.55 | 0.66 |
| Post Interest (1-4) | 2.66 | 2.27 | 0.69 |
| Physics Identity (1-4) | 2.04 | 1.64 | 0.59 |

**SEM path model**

In relation to **RQ2,** we initially tested moderation analysis using multi-group SEM between female and male students to find if any of the relationships between the variables were different across gender. There were no group differences at the level of weak and strong measurement invariance at the level of regression coefficients, so we proceeded with mediation analysis.

Then, in relation to **RQ3,** Fig. 2 shows that gender only directly predicts pre self-efficacy, pre interest, sense of belonging, and peer interaction. In particular, Fig. 2 shows that gender does not directly predict any of the outcomes. We used mediation analysis to understand the extent to which gender differences in students' outcomes at the end of the introductory physics courses (self-efficacy or S.E., interest, and physics identity) were mediated by differences in students' initial self-efficacy, interest, and perception of the learning environment (perceived recognition, peer interaction, and belonging) in the course. The model is shown in FIG. 2.

In relation to **RQ4,** we found that interest at the end of physics 1 was mainly predicted by interest at the beginning of the class (regression coefficient $\beta = 0.72$) with smaller effects from peer interaction ($\beta = 0.24$) and perceived recognition ($\beta = 0.21$). Although post interest appears to be strongly correlated with pre interest, it does not mean that interest cannot be changed throughout the class via evidence-based intentional design. For example, students' interest could increase through the learning environment from the positive contributions from peer interaction and perceived recognition. One possible way to increase students' interest in physics is to provide opportunities in class that relate to students' interests and career paths, so students can see how they could use physics in their careers. Relating physics to everyday life could be another effective approach to increasing interest (Good et al., 2018)

Post self-efficacy has direct effects from pre self-efficacy ($\beta = 0.45$), belonging ($\beta = 0.49$), peer interaction ($\beta = 0.21$), and a smaller direct effect from perceived recognition ($\beta = 0.13$). Self-efficacy is important for students' persistence in the class and future careers. Since the learning environment can predict student self-efficacy, it is important for an instructor to try to improve the belonging, peer interaction, and perceived recognition of the students. For instance, instructors can positively influence the students' peer interaction with each other by providing time for the students to work together in class and making sure all voices are heard equally when discussing problems as a class.

Self-efficacy, interest, and perceived recognition influence students' physics identity directly, with perceived recognition having the largest direct effect ($\beta = 0.59$). This is consistent with past models (Hazari et al., 2010; Kalender et al., 2019a). TABLE 3 shows us that both women and men have a mean perceived recognition below the positive threshold (score of 3) and women also have lower scores than men in all constructs, including their physics identity.

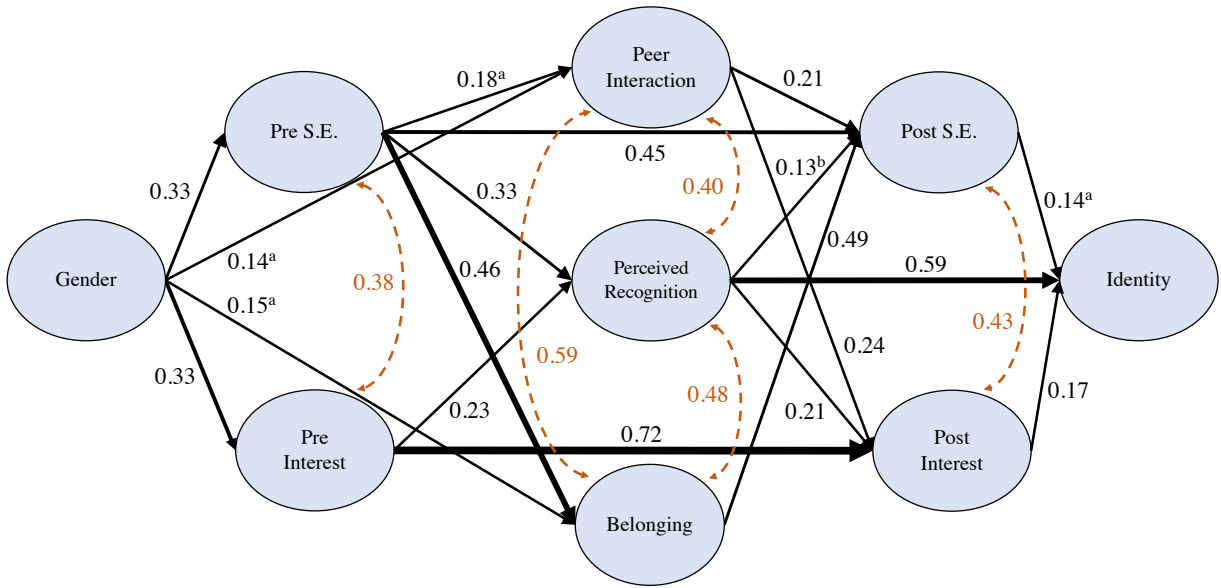

FIG. 2 RESULT OF THE SEM BETWEEN GENDER AND OUTCOMES IN PHYSICS 1 THROUGH VARIOUS MEDIATING FACTORS. THE LINE THICKNESS IS THE RELATIVE MAGNITUDE OF β VALUES. ALL *P*-VALUES ARE INDICATED BY NO SUPERSCRIPT FOR *P* <0.001, "A" FOR *P* < 0.005, AND "B" FOR *P* < 0.05 VALUES. S.E. REFERS TO STUDENTS' SELF-EFFICACY. ALL REGRESSION PATHS THAT ARE STATISTICALLY SIGNIFICANT FROM LEFT TO RIGHT BETWEEN ANY TWO CONSTRUCTS ARE SHOWN. THE DASHED LINES INDICATE COVARIANCE.

## SUMMARY, IMPLICATIONS, AND FUTURE DIRECTIONS

In this research, we find a gender gap disadvantaging women in all of the students' beliefs studied, similar to what was found in calculus-based physics courses (Kalender et al., 2019b). In addition, both men's and women's self-efficacy and interest scores dropped from the beginning to the end of the physics 1 course. Women also had a more negative perception of the inclusiveness of learning environment constructs (TABLE 3). These differences may result from structural inequities in the physics learning environment and marginalized students, e.g., women, not being adequately supported (Charleston et al., 2014; Johnson, 2012; Kellner, 2003; Morton & Parsons, 2018; Tolbert et al., 2018)

Our SEM model indicates that the perception of the inclusiveness of learning environment factors (belonging, peer interaction, and perceived recognition) explains student outcomes (self-efficacy, interest, and identity). While past research has investigated physics identity, no research has investigated factors that instructors can control in order to make their classes equitable and inclusive. Our findings suggest that it may be beneficial to implement structural changes at the classroom level that targets students' sense of belonging, peer interaction, and perceived recognition in order to improve equity and inclusion. These factors predict each other as well, so if instructors can provide support for some of the factors they can readily control (e.g., peer interaction or perceived recognition), they can make their classes more equitable, and they are likely to improve student outcomes in the process.

We note that the learning environment in many classes may not be providing the same opportunities for all students (Estrada et al., 2018; Green et al., 2019), particularly those from marginalized groups such as women in physics. Approaches to improve the inclusivity of the learning environment need to be systemic. Structural changes at the institution level require centering marginalized students in the design of curricula and pedagogies. Departments and institutions can reward instructors for supporting equity and inclusion that eliminates the problematic gender gap discussed here.

In addition, instructors can improve the learning environment in their courses by adopting culturally responsive pedagogy and providing mentoring/support for students who are underrepresented (Beckford et al., 2020). Instructors should be careful not to say that problems are "trivial", "easy" or "obvious" when students ask them for help because female students are more likely to feel disparaged (or negatively recognized) due to the stereotypes in physics. What is important for instructors to realize is that it is not their intention that matters but the impact they are having on students. Another way to improve the learning environment is through classroom interventions. Brief social-psychological classroom interventions have been shown to decrease or eliminate the gap in STEM contexts between students from marginalized and dominant groups (Binning et al., 2020; Harackiewicz et al., 2016; Walton et al., 2015; Yeager & Walton, 2011).

In summary, more should be done in the college physics classrooms to mitigate the stereotypes and past experiences women have had over their lifetime since otherwise even in a physics class in which they are not underrepresented, women will be disadvantaged. Physics instructors and TAs need to provide an equitable and inclusive learning environment that emphasizes recognizing students for making progress, allowing for positive peer interactions, and providing a space where all students can feel that they belong. From our analysis, these factors play a key role in predicting students' self-efficacy, interest, and identity in physics. It is important to note that student perception of the inclusiveness of the learning environment is not shaped only by what happens in the classroom. Student interactions with each other while they do homework, students' experiences in an instructor's or TA's office hours, interactions between students and the instructor over email, and other circumstances all contribute to students' perception of inclusiveness of learning environment and can affect students' physics identity, self-efficacy, and interest.

## ACKNOWLEDGEMENTS

This work was supported by grant NSF DUE-152457. We thank all students who participated in this research and Dr. Robert Devaty for his constructive feedback on this manuscript.